\documentclass[12pt,a4]{article} 
\newcommand{\vs}{\vspace{-0.25cm}}
\usepackage{amsmath,graphicx}
\begin{document} 
\begin{center}
  {\Large{\bf Solving the matrix exponential function for special orthogonal groups SO(n) and the exceptional G$_2$} }  

\bigskip

 Norbert Kaiser\\
\medskip
{\small Physik-Department T39, Technische Universit\"{a}t M\"{u}nchen,
   D-85748 Garching, Germany\\}

\smallskip

{\it email: nkaiser@ph.tum.de}
\end{center}

\begin{abstract}
In this work the matrix exponential function is solved analytically for the special orthogonal groups $SO(n)$ up to $n=9$. The number of occurring  $k$-th matrix powers gets limited to $0\leq k \leq n-1$ by exploiting the Cayley-Hamilton relation. The corresponding expansion coefficients can be expressed as cosine and sine functions of a vector-norm $V$ and the roots of a polynomial equation that depends on a few specific invariants. Besides the well known case of $SO(3)$, a quadratic equation needs to be solved for $n=4,5$, a cubic equation for $n=6,7$, and a quartic equation for $n=8,9$. As an interesting subgroup of $SO(7)$, the exceptional Lie group $G_2$ of dimension $14$ is constructed via the matrix exponential function through a remarkably simple constraint on an invariant, $\xi=1$. The calculation of the trace of the $SO(n)$-matrices arising from the exponential function, results in a sum of cosines of several angles, which specify the associated conjugation class as a point on a maximal torus.

\end{abstract}
\section{Introduction and summary}
For $SU(2)$-matrices the result of evaluating the matrix exponential function, $U = \exp(i \vec \tau\!\cdot\! \vec v\,) = {\bf 1} \cos|\vec v\,|  + i\vec \tau \!\cdot\! \hat v \, \sin|\vec v\,|$, with $\vec \tau=(\tau_1, \tau_2,\tau_3) $ the Pauli matrices, is well known and frequently used in effective field theories where $U$ comprises three low-energy excitations (e.g. pions) and thus serves as the basic field variable.  In a recent work \cite{matexp} the solution of the matrix exponential function has been extended to the $SU(3)$-group with eight real parameters. By employing the Cayley-Hamilton relation the required matrix powers could be reduced to the zeroth, first and second. The resulting analytical formula involved the sum over three real roots of a cubic equation, corresponding thus to the so-called irreducible case, where one employs for its solution the trisection of an angle.  When going to the special unitary group $SU(4)$ with 15 real parameters, the analytical formula involved the sum over four real roots of a quartic equation. The associated cubic resolvent equation with three positive roots belonged again to the irreducible case. By imposing the pertinent condition on $SU(4)$-matrices one could also treat the symplectic group $Sp(2)$ with ten real parameters. Since there the roots occurred as two pairs of opposite sign, this simplified the analytical formula for $Sp(2)$-matrices considerably. An outlook to the situation with semi-analytical formulas for $SU(5)$, $SU(6)$, and $Sp(3)$ has also been given in ref.\,\cite{matexp}. 

The purpose of the present work is to continue the approach of ref.\,\cite{matexp} by solving the matrix exponential function analytically for the special orthogonal groups $SO(n)$ up to $n=9$. Again, by exploiting the Cayley-Hamilton relation for the elements of the Lie algebra $so(n)$ (i.e. antisymmetric $n\times n$ matrices), the number of required matrix powers ranges from $0$ up to $n-1$. The corresponding expansion coefficients will be expressed as cosine and sine functions of a vector-norm $V= |\vec v\,|$ and the roots $y_j$ of a polynomial equation that depends on some specific invariants (the determinant and the traces of even matrix powers). Putting aside the almost trivial cases of $SO(2)$ and $SO(3)$, a quadratic equation needs to be solved for $n=4,5$, a cubic equation for $n=6,7$, and a quartic equation for $n=8,9$ (where the latter leads to a cubic resolvent equation). The exceptional Lie group $G_2\subset SO(7)$ of dimension $14$, defined as the automorphism group of the octonions, is constructed via the matrix exponential function by first deriving seven homogeneous linear relations for the $21$ parameters of a general $so(7)$ Lie algebra element. These restricting relations translate into a remarkably simple constraint on an invariant, $\xi = 1$. The calculation of the trace of the $SO(n)$-matrices arising from the exponential function,  gives as a result a simple sum of cosines of several angles, which specify the associated conjugation class as a point on a maximal torus $SO(2)\times \dots \times SO(2)$.

In the following one uses the generators $J_a, a=1,\dots, n(n-1)/2$ (i.e. basis elements of the Lie algebra $so(n)$), where the antisymmetric $n\times n$ matrix $J_a$ has exactly one entry $1$ above the diagonal that is reflected to   a $-1$ below the diagonal. These generators are normalized as tr$(J_a J_b) = -2 \delta_{ab}$. Following a gradually extended  strategy one works from the easy case $n=3$ up to $n=9$, with an insertion after $SO(7)$ that treats in detail the interesting subgroup $G_2\subset SO(7)$. In perspective, one remarks that for higher $n=10,11$ analogous semi-analytical formulas could be written down to solve the matrix exponential function, but these involve a sum over the five (positive) roots of quintic equation for which no direct algebraic solution formulas in terms of its coefficients exist.
\section{SO(3)}
Elements of the three-dimensional Lie algebra $so(3)$ are antisymmetric $3\times 3$ matrices of the form
\begin{equation}\vec J\!\cdot\! \vec v  =  \left( \begin{array} {ccc} 0 & v_1& v_2
\\ -v_1& 0& v_3\\ -v_2& -v_3& 0 \cr \end{array}\!\right), 
\end{equation} 
with the length $ V=\sqrt{v_1^2+v_2^2+v_3^2}$ of the three-component real vector $\vec v$.
The normalized matrix $\Sigma = \vec J \!\cdot\! \hat v= \vec J \!\cdot\! \vec v/V$ satisfies the relation $\Sigma^3 = -\Sigma$ and therefore all powers of $\Sigma $ can be reduced to the first two. The matrix exponential function for the special orthogonal group $SO(3)$  takes the following simple form
\begin{equation}R_3(\vec v\,)=\exp(\vec J\!\cdot\! \vec v\,) = \mathbf{1} +\sin V\, \vec J \!\cdot\! \hat v+ (1-\cos V) \, (\vec J \!\cdot\! \hat v)^2\,, \end{equation} 
where $\mathbf{1}$ denotes the $3\times 3$ unit-matrix.
The formula in eq.(2) is well known to describe a right-handed rotation in three-dimensional space about the axis $\vec n =(-v_3,v_2,-v_1)/V$ with an angle
$\varphi =V$. This angle follows also directly from the trace, tr$R_3(\vec v\,)=1 +2\cos V$. Note that $\varphi=V $ parametrizes at fixed $\hat v$ a so-called maximal torus $SO(2)=\mathbf{S}^1$ (circle line) in $SO(3) = \mathbf{RP}^3$ (three-dimensional real projective space \cite{liegroup}). For comparison the almost trivial result for $SO(2)$-rotations in a plane reads:
\begin{equation} R_2(v)= \exp( I v) = \mathbf{1}  \cos v +  I  \sin v\,, \qquad I = \left( \begin{array} {cc} 0 & 1\\ -1& 0 \cr \end{array}\!\right), \quad I^2 = - \mathbf{1}  \end{equation}   
with the trace, tr$R_2(v) = 2\cos v$, of this $2\times 2$ rotation matrix.
\section{SO(4)}
Elements of the six-dimensional Lie algebra $so(4)$ are antisymmetric $4\times 4$ matrices of the form
\begin{equation}\vec J\!\cdot\! \vec v  =  \left( \begin{array} {cccc} 0 & v_1& v_2&v_3
\\ -v_1& 0& v_4&v_5\\ -v_2& -v_4& 0 &v_6 \\ -v_3& -v_5& -v_6& 0\cr \end{array}\!\right), 
\end{equation} 
with the corresponding length $ V=\sqrt{v_1^2+v_2^2+v_3^2+v_4^2 +v_5^2+v_6^2}$ of the six-component real vector $\vec v$. For any antisymmetric matrix the traces of its odd powers vanish altogether. Based on the relation tr$(\Sigma^2)=-2$ the characteristic polynomial \cite{koecher} of the normalized matrix $\Sigma = \vec J \!\cdot\! \hat v=\vec J \!\cdot\! \vec v/V$ is given by
\begin{equation}P_4(x) = x^4+x^2+\eta\,,\end{equation} 
with the invariant 
\begin{equation}\eta ={\rm det}(\vec J \!\cdot\! \hat v) =\Big({v_1v_6-v_2v_5+v_3v_4 \over v_1^2+v_2^2+v_3^2+v_4^2 +v_5^2+v_6^2}\Big)^2\geq 0  \,.\end{equation}
The four roots of $P_4(x)=0$ are the eigenvalues of $\Sigma$ and these must be purely imaginary, since $ \Sigma= -\Sigma^\dagger$ is antihermitean (actually it is real antisymmetric). This fact about the roots implies for the solutions $-x_\pm^2$  of the intermediate quadratic equation (in $-x^2$) the inequality
\begin{equation} -2x^2_\pm =1 \pm \sqrt{1-4\eta} \geq 0\,,\end{equation}
from which one deduces the allowed range $0\leq \eta\leq 1/4$ for the determinant $\eta$. As a consequence of the Cayley-Hamilton relation $\Sigma^4 = -\Sigma^2 -\eta \mathbf{1}$ the number of independent matrix powers that get produced by the exponential series is limited to the first three (including the $4\times 4$ unit-matrix $\mathbf{1}$).  
Starting at order $n$ with $\Sigma^n =\alpha_n \, \mathbf{1}+\beta_n\,  \Sigma +\gamma_n\,  \Sigma^2+\delta_n\,  \Sigma^3 $ and multiplying with $\Sigma$, one obtains via the mentioned relation the expansion coefficients at order $n+1$. The resulting linear recursion reads in matrix-vector notation
\begin{equation} \left(\!\begin{array} {c}\alpha_{n+1} \\ \beta_{n+1} \\ \gamma_{n+1} \\ \delta_{n+1} \cr\end{array}\!\right) = M_4  \left(\!\begin{array} {c}\alpha_n \\ \beta_n \\ \gamma_n \\ \delta_n \cr\end{array}\!\right), \quad  \quad M_4 =  \left( \begin{array} {cccc} 0 & 0& 0&-\eta 
\\ 1& 0& 0& 0\\ 0&1& 0&-1 \\ 0& 0& 1&0\cr \end{array}\!\right), \end{equation} 
and the initial values are $\alpha_0=1, \beta_0=0, \gamma_0=0, \delta_0 =0$.
By diagonalization the exponential series $\exp(V M_4 )=\sum_{k=0}^\infty  (V M_4)^k/k!$ can be solved and after vector multiplication with $(1,0,0,0 )$ from the right and $( \mathbf{1}, \Sigma,\Sigma^2,\Sigma^3)$ from the left, one ends up with the following analytical formula for $SO(4)$ rotation matrices:
\begin{eqnarray} R_4(\vec v\,) =\exp(\vec J \!\cdot \!\vec v\,) &=& {1\over 1-2z} \bigg\{ \Big[ (1-z)\cos\big(V \sqrt{z}\,\big)- z \cos\big(V \sqrt{1-z}\,\big)\Big] \mathbf{1}  \nonumber \\ && +\Big[{1-z\over \sqrt{z} }\sin\big(V \sqrt{z}\,\big)- {z\over \sqrt{1-z}} \sin\big( V \sqrt{1-z}\,\big)\Big] \vec J \!\cdot \!\hat v
\nonumber \\ && +\Big[\cos\big(V \sqrt{z}\,\big)- \cos\big( V \sqrt{1-z}\,\big)\Big] (\vec J \!\cdot \!\hat v)^2
\nonumber \\ && +\Big[{1\over \sqrt{z} }\sin\big(V \sqrt{z}\,\big)- {1\over \sqrt{1-z}} \sin\big( V \sqrt{1-z}\,\big)\Big] (\vec J \!\cdot \!\hat v)^3\bigg\}\,.
\end{eqnarray}
The auxiliary parameter $z$ is introduced via the relation 
$\eta =z(1-z)$ to the determinant $\eta$. Since eq.(9) is invariant under the substitution $z\to 1-z$, one can restrict the  values of $z$ to the interval $0\leq z \leq 1/2$, taking the solution $z= (1-\sqrt{1-4\eta}\,)/2$. Note that the four expansion coefficients in eq.(9) depend only on $V$ and $z$. Calculating  the trace of the special orthogonal $4\times 4$ matrix, tr$R_4(\vec v\,)=2\cos(V \sqrt{z}\,)+ 2\cos( V \sqrt{1-z}\,)$, reveals that the two involved rotation angles are $\varphi_1 = V\sqrt{1-z}$ and $\varphi_2 = V\sqrt{z}$, which satisfy the conditions $\varphi_1\geq \varphi_2$, $\varphi_1^2+\varphi_2^2 =V^2$ and $\varphi_1\varphi_2 =\sqrt{\eta}\,V^2$. Both angles specify to which element on a maximal torus $SO(2)\times SO(2) = \mathbf{S}^1 \times \mathbf{S}^1$ in $SO(4) = \mathbf{S}^3 \times \mathbf{RP}^3$ the given rotation matrix $R_4(\vec v\,)$ is related by conjugation \cite{liegroup}. Note that a conventional maximal torus  in $SO(4)$ consists of independent rotations that take place in the $x_1x_2$-plane and the $x_3x_4$-plane  of four-dimensional space.
\section{SO(5)}
Elements of the ten-dimensional Lie algebra $so(5)$ are antisymmetric $5\times 5$ matrices of the form
\begin{equation}\vec J\!\cdot\! \vec v  =  \left( \begin{array} {ccccc} 0 & v_1& v_2&v_3&v_4
\\ -v_1& 0& v_5&v_6&v_7\\ -v_2& -v_5& 0 &v_8&v_9 \\ -v_3& -v_6& -v_8& 0&v_{10}\\ -v_4& -v_7& -v_9&-v_{10}&0 \cr \end{array}\!\right), 
\end{equation} 
with the corresponding norm $ V=\sqrt{v_1^2+\dots +v_{10}^2}$ of the ten-component real vector $\vec v$. Since the determinant of $\vec J\!\cdot\! \vec v$ vanishes (in all odd dimensions), det$(\vec J\!\cdot\! \vec v\,)=0$, one has as a new invariant for $so(5)$-matrices  the trace of the fourth power, $\xi = \text{ tr}(\Sigma^4)$, where $\Sigma =  \vec J\!\cdot\! \hat v=\vec J\!\cdot\! \vec v/V$. The invariant $\xi$ enters the characteristic polynomial \cite{koecher} of degree five:
\begin{equation} P_5(x) = x^5+x^3+x\Big( {1\over 2}-{\xi\over 4}\Big)\,,\end{equation}
Again all its roots (including zero) must be purely imaginary. This fact implies for the solutions $-x_\pm^2$  of the intermediate quadratic equation (in $-x^2$) the inequality
 \begin{equation}-2x^2_\pm = 1 \pm \sqrt{\xi-1} \geq 0\,,\end{equation}
 from which one deduces the allowed range of $\xi$ as the interval $1\leq \xi\leq 2$.  Based on the Cayley-Hamilton relation $\Sigma^5 = -\Sigma^3 +\Sigma( \xi-2)/4$  one constructs the $5\times 5$ iteration matrix as
 \begin{equation}M_5  =  \left( \begin{array} {ccccc} 0 & 0& 0&0 &0 
\\ 1& 0& 0& 0 &\xi/4\!-\!1/2\\ 0& 1 & 0 & 0& 0 \\ 0& 0& 1& 0&-1\\ 0& 0& 0& 1& 0\cr \end{array}\!\right), \end{equation} 
with which one can solve $\exp(V M_5)$ and multiply with $(1,0,0,0,0)$. 
It is furthermore advantageous to parametrize $\xi= 1+\cos^22\theta$ in terms of an angle  $\theta$, that is taken from the interval $0\leq \theta \leq \pi/4$, namely $\theta = {1\over 2}\arccos\sqrt{\xi-1}$,  Putting all pieces together, the solution of the matrix exponential function for the $SO(5)$ rotation group  reads 
\begin{eqnarray} R_5(\vec v\,)=\exp(\vec J \!\cdot \!\vec v\,) &=&\mathbf{1} +  {1\over \cos2\theta} \bigg\{\bigg[ {\cos^2\theta \over \sin\theta} \sin(V \!\sin\theta) -{\sin^2\theta \over \cos\theta}\sin(V\! \cos\theta) \bigg] \vec J \!\cdot \!\hat v \nonumber \\ &&
+ \bigg[ {1\over \sin^2\theta} -{1\over \cos^2\theta} +\tan^2\theta \cos(V \!\cos\theta) -\cot^2\theta \cos(V \!\sin\theta)\bigg] (\vec J \!\cdot \!\hat v)^2  \nonumber \\ &&+ \bigg[ {\sin(V \!\sin\theta)\over \sin\theta} -{\sin(V \!\cos\theta)\over \cos\theta} \bigg](\vec J \!\cdot \!\hat v)^3 \nonumber \\ && + \bigg[ {1-\cos(V\!\sin\theta) \over \sin^2\theta}- {1-\cos(V\!\cos\theta) \over \cos^2\theta}\bigg] (\vec J \!\cdot \!\hat v)^4 \bigg\}\,, \end{eqnarray} 
where $\mathbf{1} $ denotes the $5\times 5$ unit-matrix.
Note again that the four expansion coefficients depend only on $V$ and $\theta$. The occurring trigonometric functions of $\theta$ are related to the invariant $\xi\in[1,2]$ by the relations
\begin{eqnarray} && \cos2\theta = \sqrt{\xi-1}\,, \quad \sin\theta = {\sqrt{1-\sqrt{\xi-1}}\over \sqrt{2}} \,, \quad \cos\theta = {\sqrt{1+\sqrt{\xi-1}}\over \sqrt{2}}\,, \nonumber \\ && \tan\theta = {1 -\sqrt{\xi-1} \over \sqrt{2-\xi}}\,, \quad \cot\theta = {1 +\sqrt{\xi-1} \over \sqrt{2-\xi}}\,.   \end{eqnarray} 
By calculating the trace of the special orthogonal matrix, tr$R_5(\vec v\,)=1+2\cos(V\! \cos\theta)+ 2\cos( V\! \sin\theta)$,  one recognizes that the two involved rotation angles are $\varphi_1 = V\cos\theta$ and $\varphi_2 = V\sin\theta $, subject to the conditions $\varphi_1\geq \varphi_2$, $\varphi_1^2+\varphi_2^2 =V^2$ and $\varphi_1\varphi_2 =\sqrt{2-\xi}\,V^2/2$. Again, these two angles specify the conjugation class \cite{liegroup} of $R_5(\vec v\,)$ as a point on the maximal torus  $SO(2)\times SO(2) = \mathbf{S}^1 \times 
\mathbf{S}^1$ in $SO(5)$. By convention such a maximal torus consists of independent rotations in the $x_1x_2$-plane and  the $x_3x_4$-plane  of five-dimensional space.

\section{SO(6)}

\begin{figure}\centering
\includegraphics[width=0.7\textwidth]{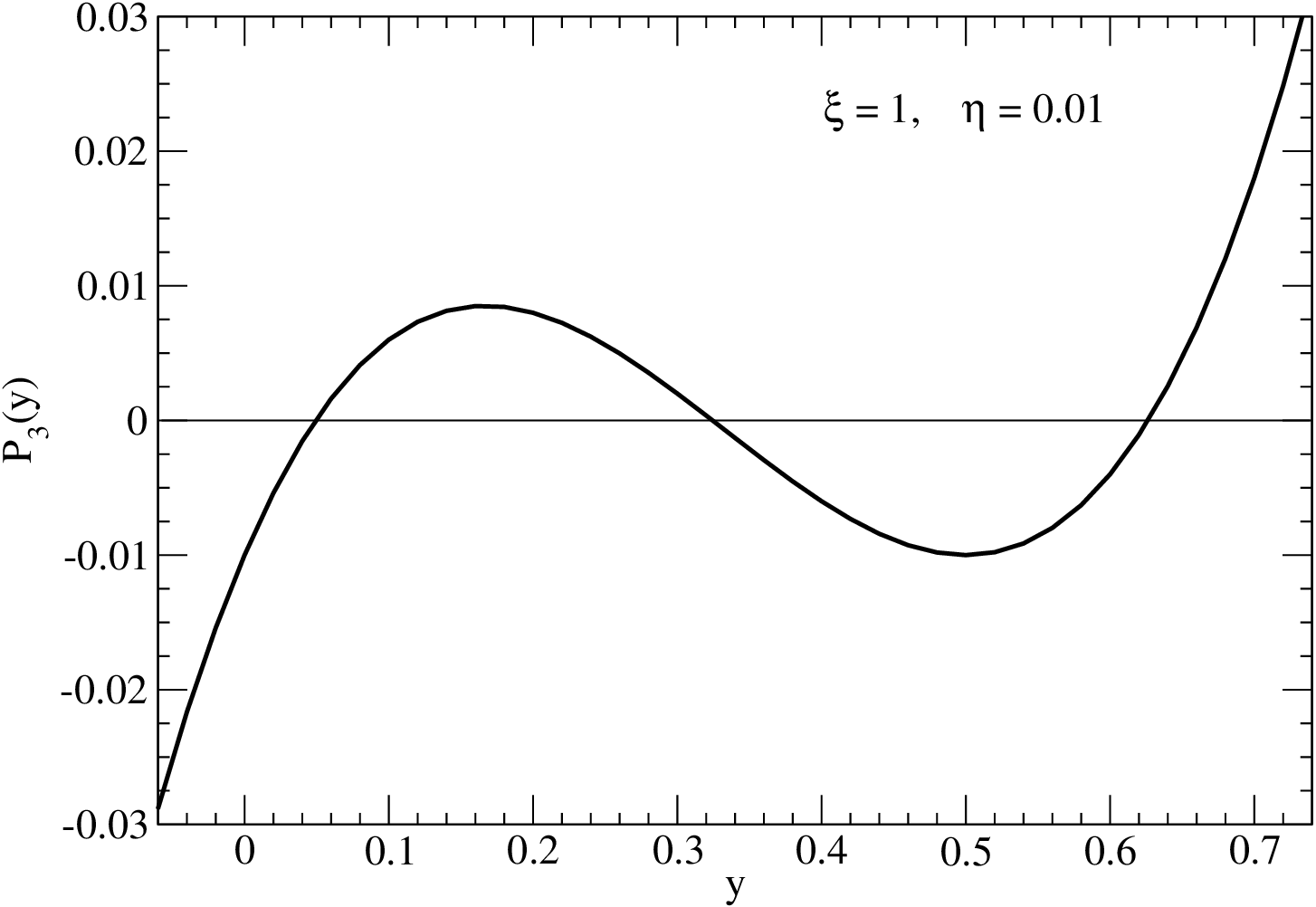}
\caption{Generic behavior of the cubic polynomial $\tilde P_3(y) = y^3-y^2+y(1/2-\xi/4) -\eta$.} 
\end{figure}
Elements of the 15-dimensional Lie algebra $so(6)$ are antisymmetric $6\times 6$ matrices of the form
\begin{equation}\vec J\!\cdot\! \vec v  =  \left( \begin{array} {cccccc} 0 & v_1& v_2&v_3&v_4&v_5
\\ -v_1& 0& v_6&v_7 &v_8&v_9\\ -v_2& -v_6& 0 &v_{10}&v_{11}&v_{12} \\ -v_3& -v_7& -v_{10}& 0&v_{13}& v_{14}\\ -v_4& -v_8& -v_{11}&-v_{13}& 0&v_{15}\\ -v_5& -v_9& -v_{12}&-v_{14}& -v_{15}& 0\cr \end{array}\!\right), 
\end{equation} 
with the corresponding norm $ V=\sqrt{v_1^2+\dots +v_{15}^2}$ of the 15-component real vector $\vec v$. The characteristic polynomial \cite{koecher} of $\Sigma = \vec J\!\cdot \!\hat v$ is of degree six and it reads
 \begin{equation} P_6(x) = x^6+x^4+x^2\Big({1\over 2}-{\xi\over 4}\Big) +\eta\,,
 \end{equation}
 with coefficients $\xi= \text{tr}(\vec J\!\cdot \!\hat v)^4$ and $\eta = 
 \text{det}(\vec J\!\cdot \!\hat v)$.  The substitution $y=-x^2$ leads to a  cubic polynomial (of half degree)
 \begin{equation} \tilde P_3(y) = y^3-y^2+y\Big({1\over 2}-{\xi\over 4}\Big) -\eta\,,
 \end{equation}
 whoose three roots $y_1, y_2, y_3\geq 0$ all have to be positive, since those of $P_6(x)=0$ are the purely imaginary eigenvalues of the antihermitean $6\times 6$ matrix $\Sigma$. Fig.\,1 shows the generic behavior of such a cubic polynomial. From $\tilde P_3(0)=-\eta \leq0$ one learns first $\eta\geq 0$, and fact that the positions of the local minimum and maximum at $y_{\pm}$ lie on the positive $y$-axis, leads to the inequality
    \begin{equation} 6 y_\pm = 2 \pm \sqrt{3\xi-2} \geq0\,,
 \end{equation} 
 from which one deduces the allowed range $2/3\leq \xi\leq 2$ for the invariant $\xi$. Moreover, the product $\tilde P(y_+)\tilde P(y_-)\leq 0$ is negative, and this condition gives rise to a further inequality:
  \begin{equation} (3\xi-2)^3 \geq (9\xi+108\eta -10)^2 \,.
 \end{equation}

\begin{figure}[ht]\centering
\includegraphics[width=0.7\textwidth]{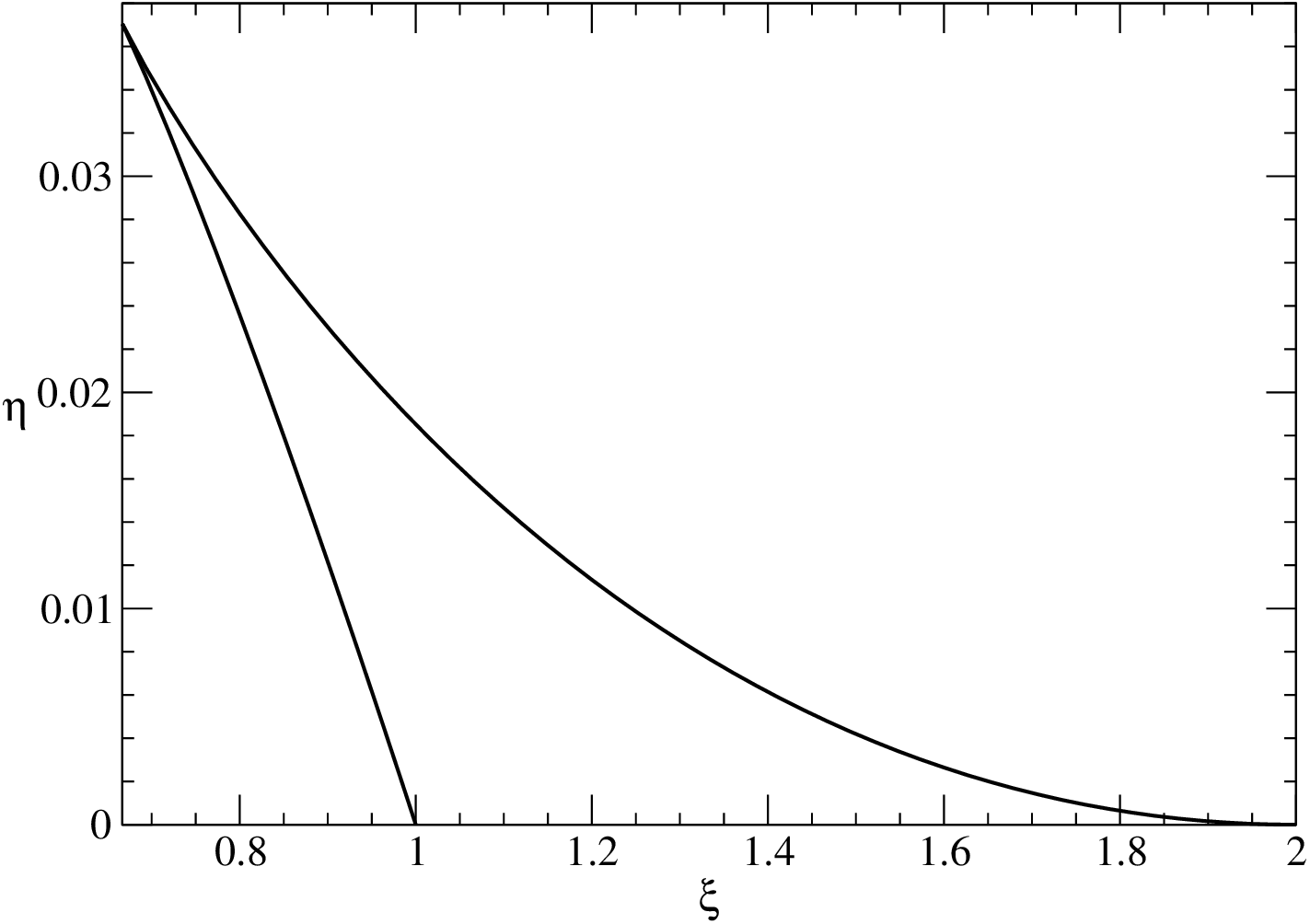}
\caption{The allowed values of the invariants  $\xi$ and $\eta$  for $so(6)$ lie inside the bounded region. The enclosed area in the $\xi\eta$-plane amounts to $1/120$, and the circumference of the tricorn measures $2.6695$.} 
\end{figure}

The resulting allowed range for the invariants $\xi$ and $\eta$ is the bounded region shown in Fig.\,2, from which one deduces also the maximal value $\eta_\text{max} = 1/27$. 
With three positive real roots the cubic polynomial $\tilde P_3(y)$ corresponds to the so-called irreducible case, where the problem is effectively solved by the trisection of an angle. The ansatz $y_1 = (1+\sqrt{3\xi-2}\cos\psi)/3$ leads to a determining equation for $\cos 3\psi$, that is immediately solved by   
 \begin{equation} \psi = {1\over 3} \arccos{9\xi+108\eta -10 \over (3\xi-2)^{3/2}} \,,\end{equation} 
with $\psi\in[0,\pi/3]$. Note that the inequality derived in eq.(20) guarantees that the argument of the arc-cosine function lies between $-1$ and $1$.  The other two roots are given by
 \begin{equation} y_{2,3} ={1\over 6}\big[ 2 +\sqrt{3\xi-2}\big(\pm \sqrt{3} \sin\psi-\cos\psi\big)\big]\,, \end{equation} 
 and with this assignment the three roots are ordered as $y_1\geq  y_2\geq y_3\geq 0$.
 
 Based on the Cayley-Hamilton relation $\Sigma^6 = -\Sigma^4 +\Sigma^2( \xi-2)/4-\eta \mathbf{1}$  one constructs the $6\times 6$ iteration matrix as
 \begin{equation}M_6  =  \left( \begin{array} {cccccc} 0 & 0& 0&0 &0&-\eta \\ 1 & 0& 0&0 &0& 0 \\ 0& 1& 0& 0& 0 &\xi/4\!-\!1/2\\ 0& 0& 1 & 0 & 0& 0 \\0 & 0& 0& 1& 0&-1\\ 0& 0& 0& 0& 1& 0\cr \end{array}\!\right), \end{equation} 
with which one can solve $\exp(V M_6)$ and multiply with $(1,0,0,0,0,0)$. 
One ends up with the following quasi-analytical formula for the matrix exponential function $so(6)\to SO(6)$: 
\begin{eqnarray} R_6(\vec v\,)=\exp(\vec J \!\cdot \!\vec v\,) &=&\sum_{j=1}^3 {1\over 3 y_j^2-2y_j+1/2-\xi/4} \Big\{ {\eta \over y_j} \mathbf{1} +(1-y_j) (\vec J \!\cdot \!\hat v)^2 +(\vec J \!\cdot \!\hat v)^4 \Big\}\nonumber \\ && \qquad  \times  \Big\{ \cos\big(V\!\sqrt{y_j}\big) \mathbf{1}+ {1\over \sqrt{y_j}} \sin\big(V\!\sqrt{y_j}\big)  \vec J \!\cdot \!\hat v\Big\} \,,
\end{eqnarray} 
where the sum goes over the three positive roots of the cubic equation $\tilde P_3(y)=0$. Note that the denominator of the prefactor is the derivative $\tilde P_3'(y_j)$, and one finds a remarkable factorization for the terms with (up to fifth) powers of $\vec J \!\cdot \!\hat v$. When taking  the trace of the special orthogonal matrix $R_6(\vec v\,)$ one gets 
\begin{equation}\text{tr} R_6(\vec v\,)=\sum_{j=1}^3 {\cos\big(V\!\sqrt{y_j}\big)\over 3 y_j^2-2y_j+1/2-\xi/4}\Big\{ {6 \eta \over y_j} +2(y_j-1) +\xi\Big\} =  2\sum_{j=1}^3 \cos\big(V\!\sqrt{y_j}\big)\,, \end{equation}
where the final expression is obtained by eliminating $\eta$ in favor of the root $y_j$, using its determining equation $\tilde P_3(y_j)=0$. The three angles
$\varphi_j = V\!\sqrt{y_j}$  satisfy the conditions $\varphi_1  \geq \varphi_2 \geq \varphi_3$, $ \varphi_1^2+\varphi_2^2+\varphi_3^2=V^2$ (due to the root-sum $y_1+y_2+y_3=1$), $\varphi_1 \varphi_2\varphi_3 = \sqrt{\eta}\,V^3$.  In $SO(6)$ the  maximal torus  $SO(2)\times SO(2)\times SO(2) = \mathbf{S}^1 \times \mathbf{S}^1 \times \mathbf{S}^1$ is three-dimensional and  by convention the three independent rotations occur in the $x_1x_2$-plane, $x_3x_4$-plane, and $x_5x_6$-plane of six-dimensional space. 

\section{SO(7)}
Elements of the 21-dimensional Lie algebra $so(7)$ are antisymmetric $7\times 7$ matrices of the form
\begin{equation}\vec J\!\cdot\! \vec v  =  \left( \begin{array} {ccccccc} 0 & v_1& v_2&v_3&v_4&v_5&v_6
\\ -v_1& 0& v_7&v_8 &v_9&v_{10}& v_{11}\\ -v_2& -v_7& 0 &v_{12}&v_{13}&v_{14} &v_{15} \\ -v_3& -v_8& -v_{12}& 0&v_{16}& v_{17}& v_{18}\\ -v_4& -v_9& -v_{13}&-v_{16}& 0&v_{19}& v_{20}\\ -v_5& -v_{10}& -v_{14}&-v_{17}& -v_{19}& 0& v_{21}\\ -v_6& -v_{11}& -v_{15}&-v_{18}& -v_{20}& - v_{21}&0\cr \end{array}\!\right), 
\end{equation} 
with the corresponding norm $ V=\sqrt{v_1^2+\dots +v_{21}^2}$ of the 21-component real vector $\vec v$. The characteristic polynomial \cite{koecher} of $\Sigma = \vec J\!\cdot \!\hat v$ is of degree seven and it reads
 \begin{equation} P_7(x) = x^7+x^5+x^3\Big({1\over 2}-{\xi\over 4}\Big) +x\Big({1-\zeta\over 6}-{\xi\over 4} \Big) \,,
 \end{equation}
 with in addition to $\xi= \text{tr}(\vec J\!\cdot \!\hat v)^4$ a new invariant $\zeta= \text{tr}(\vec J\!\cdot \!\hat v)^6$. Besides the trivial root $x=0$ of $P_7(x)$ the other purely imaginary ones are found via the substitution $y=-x^2$ from cubic polynomial
 \begin{equation} \tilde P_3(y) = y^3-y^2+y\Big({1\over 2}-{\xi\over 4}\Big) +{\zeta-1 \over 6}+{\xi\over 4}\,. 
 \end{equation}
 It has the same form as the cubic polynomial in eq.(18) relevant for $so(6)$ after identifying the  constant term with $-\eta$. Although det$(\vec J\!\cdot \!\hat v)=0$ in $so(7)$, this connection  motivates to introduce the auxiliary  parameter
 \begin{equation} \eta_7 = {1-\zeta \over 6} -{\xi \over 4}\geq 0\,. \end{equation}
 The steps to construct the three positive roots $y_1, y_2, y_3$ of $\tilde P_3(y)=0$ are the same as in the previous section about $so(6)$. One just replaces $\eta$ by $\eta_7$ and gets for the trisected angle $\psi$ the modified expression
  \begin{equation} \psi = {1\over 3} \arccos{8-18(\xi+\zeta) \over (3\xi-2)^{3/2}}\,,
  \end{equation}
and the inequality in eq.(20) turns into $(3\xi-2)^3 \geq 4[4-9(\xi+\zeta)]^2$. The resulting allowed range for the invariants $\xi$ and $\zeta$ is the thin bounded region shown in Fig.\,3, from which one deduces also the extremal values $\zeta_\text{max} = -2/9$ and $\zeta_\text{min} = -2$. Note that this bounded region is obtained from the one shown in Fig.\,2 by the shear-transformation $\zeta =1 - 6\eta  -3\xi/2$ from the  $\xi\eta$-plane to the $\xi\zeta$-plane. 

\begin{figure}[ht]\centering
\includegraphics[width=0.7\textwidth]{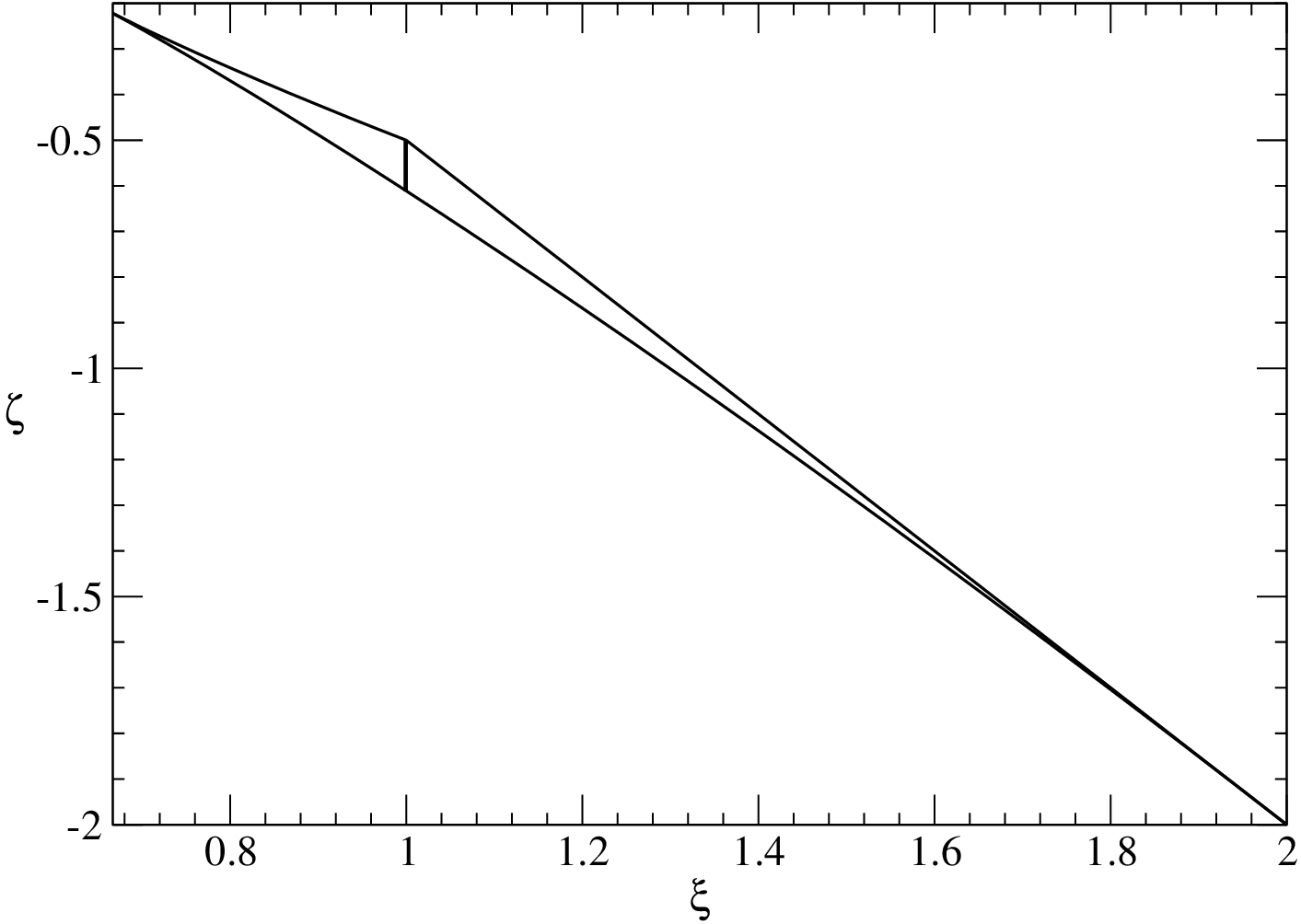}
\caption{The allowed values of the invariants  $\xi$ and $\zeta$ for $so(7)$ lie inside the thin bounded region. The enclosed area in the $\xi\zeta$-plane amounts to $1/20$, and the circumference of the tricorn measures $4.4612$. The short vertical line at $\xi=1$ of length $1/9$ corresponds to the subgroup $G_2\subset SO(7)$.}  
\end{figure}

Based on the Cayley-Hamilton relation $\Sigma^7 = -\Sigma^5 +\Sigma^3( \xi-2)/4-\eta_7 \Sigma$  one constructs the $7\times 7$ iteration matrix as
 \begin{equation}M_7  =  \left( \begin{array} {ccccccc}0 & 0& 0&0 &0& 0& 0 \\ 1& 0 & 0& 0&0 &0&-\eta_7 \\ 0& 1 & 0& 0&0 &0& 0 \\ 0& 0& 1& 0& 0& 0 &\xi/4\!-\!1/2\\0 &0& 0& 1 & 0 & 0& 0 
 \\ 0 &0 & 0& 0& 1& 0&-1\\ 0 &0& 0& 0& 0& 1& 0\cr \end{array}\!\right), \end{equation} 
with which one can solve $\exp(V M_7)$ and multiply with $(1,0,0,0,0,0,0)$. 
One ends up with the following quasi-analytical formula for the matrix exponential function $so(7) \to SO(7)$: 

\begin{eqnarray} R_7(\vec v\,)=\exp(\vec J \!\cdot \!\vec v\,) &=& \mathbf{1}+{1\over \eta_7} \Big\{ \Big({1\over 2} -{\xi\over 4}\Big) (\vec J \!\cdot \!\hat v)^2 +(\vec J \!\cdot \!\hat v)^4 +(\vec J \!\cdot \!\hat v)^6\Big\}  \nonumber \\ && +2 \sum_{j=1}^3
\Big[ 7y_j^3-5y_j^2+{3\over 4}(2-\xi) y_j -\eta_7\Big]^{-1}  \Big\{ {\eta_7 \over y_j} \mathbf{1} +(1-y_j) (\vec J \!\cdot \!\hat v)^2 +(\vec J \!\cdot \!\hat v)^4 \Big\}\nonumber \\ &&  \qquad\quad  \times 
\Big\{ \sqrt{y_j} \sin\big(V\!\sqrt{y_j}\big)  \vec J \!\cdot \!\hat v -\cos\big(V\!\sqrt{y_j}\big) (\vec J \!\cdot \!\hat v)^2  \Big\} \,,
\end{eqnarray} 
where the sum goes over the three positive roots of the cubic equation $\tilde P_3(y)=0$ in eq.(28) and $\eta_7=(1-\zeta)/6-\xi/4$ is an auxiliary parameter. Again one finds a remarkable factorization for the terms with (up to sixth) powers of $\vec J \!\cdot \!\hat v$. Note also the extra terms outside the sum in addition to the $7\times 7$ unit-matrix $\mathbf{1}$. Taking the trace of the special orthogonal matrix $\text{tr} R_7(\vec v\,)$ one finds the result
\begin{equation}\text{tr} R_7(\vec v\,)=1+ 2\sum_{j=1}^3 \cos\big(V\!\sqrt{y_j}\big)\,, \end{equation}
where this short expression is obtained by first using $\zeta=1-3\xi/2-6 \eta_7$,  and then eliminating $\eta_7$ in favor of the root $y_j$ through the equation $\tilde P_3(y_j)=0$. The three angles $\varphi_j = V\!\sqrt{y_j}$ appearing in eq.(33) satisfy the conditions $\varphi_1  \geq \varphi_2 \geq \varphi_3$, $ \varphi_1^2+\varphi_2^2+\varphi_3^2=V^2$, $\varphi_1 \varphi_2\varphi_3 = \sqrt{\eta_7}\,V^3$ and they parametrize a three-dimensional maximal torus  $SO(2)\times SO(2)\times SO(2) = \mathbf{S}^1 \times \mathbf{S}^1 \times \mathbf{S}^1$ in $SO(7)$. By convention the three independent rotations take place in the $x_1x_2$-plane, $x_3x_4$-plane,  and $x_5x_6$-plane of seven-dimensional space.  

\section{Exceptional Lie group $G_2$}
The special orthogonal group $SO(7)$  of dimension 21 (and rank 3) has an interesting subgroup, namely the exceptional Lie group $G_2$ of dimension 14 (and rank 2). From the point of view of geometrical symmetries, $G_2$ is interpreted as the automorphism group
of the octonions $\mathbf{O}$, which form an eight-dimensional real division algebra spanned by $1$ and seven imaginary units $\mathbf{i}_j, j=1,\dots, 7$. The multiplication $*$ of two imaginary units is anticommutative
  \begin{equation} \mathbf{i}_j\!*\!\mathbf{i}_k + \mathbf{i}_k\!*\!\mathbf{i}_j = -2\delta_{jk} 1\,,
  \end{equation}
and non-associative for higher products. The multiplication rule for two different imaginary units  reads
 \begin{equation} \mathbf{i}_j\!*\!\mathbf{i}_k = \sum_{l=1}^7 f_{jkl} \,\mathbf{i}_l\,,\qquad j\neq k\neq l \,,\end{equation}
with totally antisymmetric structure constants $f_{jkl}$. Among 480 equivalent realizations a possible choice is to set $f_{jkl}=1$ for the cyclic index-combinations $ijk = 123, 145, 176, 246, 257, 347, 365$. 

An automorphism of $\mathbf{O}$ is defined as a linear transformation to seven new imaginary units: 
\begin{equation} \mathbf{i}'_j = \sum_{m=1}^7 S_{jm} \mathbf{i}_m \,,\end{equation}
which leaves the entire multiplication table invariant. From anticommutativity $\mathbf{i}'_j\!*\!\mathbf{i}'_k + \mathbf{i}'_k\!*\!\mathbf{i}'_j = -2\delta_{jk} 1$ one deduces $\sum_{m=1}^7 S_{jm}S_{km} = \delta_{jk}$, thus $S$ is a $7\times 7$ orthogonal matrix. Since $S=-\mathbf{1}$ changes the sign in eq.(35)  and $O(7) = SO(7)\times \{-\mathbf{1},\mathbf{1}\}$ as a group, one arrives at the necessary condition $S\in SO(7)$. From the  multiplication rule involving the structure constant $f_{jkl}$ one derives the set of coupled cubic equations
 \begin{equation} \sum_{m,n,o=1}^7 S_{jm} S_{kn} S_{lo} f_{mno}= f_{jkl} \,,\end{equation}
 which state that $G_2$ is the invariance group of an alternating  trilinear form in seven variables. Note that orthogonality has been used to bring all three $S$-matrices onto the left hand side of eq.(37). In order to deduce the implication of eq.(37) on the Lie algebra elements one considers infinitesimal transformations $S_{jm} = \delta_{jm} + \epsilon\, T_{jm}+\dots $ with $T = \vec J\!\cdot\!\vec v  \in so(7)$. The resulting linear equations read
\begin{equation} \sum_{m=1}^7\big( T_{jm}f_{mkl} + T_{km}f_{mlj} + T_{lm}f_{mjk}\big)  =0 \,.\end{equation}
By analyzing these 343 equations in packages of $49$,  one gets from $l=1$ six linear relations for the components of $\vec v$, from $l=2$ one further linear relation, but no further constraints from $l=3,4,5,6,7$. The resulting seven homogeneous linear relations for the $21$ components of $\vec v$ read
\begin{eqnarray}&& v_{12} = v_5-v_9\,, \qquad v_{13} = v_6+v_8\,, \qquad v_{14} = v_{11}-v_3\,, \qquad v_{15} = -v_4-v_{10}\,, \nonumber \\ &&  v_{19} = v_1+v_{18}\,, \qquad v_{20} = v_2-v_{17}\,, \qquad v_{21} = v_7+v_{16}\,,\end{eqnarray} 
and these specify how the 14-dimensional exceptional Lie algebra $\mathbf{g}_2$
 can be projected out of the 21-dimensional Lie algebra $so(7)$. After introducing 14 new parameters $w_1, \dots, w_{14}$ via
\begin{eqnarray}&& w_{1,8} = v_1\pm v_{18}\,, \qquad w_{2,9} = v_2\mp v_{17}\,, \qquad w_{3,10} = v_3\mp v_{11}\,, \qquad w_{4,11} = v_4\pm v_{10}\,, \nonumber \\ &&  w_{5,12} = v_5\mp v_9\,, \qquad w_{6,13} = v_6\pm v_8\,, \qquad w_{7,14} = v_7\pm v_{16}\,,\end{eqnarray} 
 the length square of the yet 21-component vector $\vec v $ becomes a sum of 14 squares:
 \begin{equation}V^2 = {3\over 2}(w_1^2+w_2^2+w_3^2+w_4^2+w_5^2+w_6^2+w_7^2)
   + {1\over 2} (w_8^2+w_9^2+w_{10}^2+w_{11}^2+w_{12}^2+w_{13}^2+w_{14}^2) \,.\end{equation}
More interesting is the effect of the seven linear relations in eq.(39) on the invariants $\xi$ and $\zeta$. The explicit calculation gives tr$(\vec J\!\cdot\!\vec v\,)^4 = (\vec v\!\cdot\!\vec v\,)^2$, which translates into the remarkable constraint
\begin{equation}\xi = \text{tr}(\vec J\!\cdot\!\hat v\,)^4  = 1\,, \qquad \text{for}\quad \mathbf{g}_2\subset so(7)\,, \end{equation}
whereas the other invariant $\zeta= \text{tr}(\vec J\!\cdot\!\hat v\,)^6$ is confined to the small interval $-11/18\leq \zeta\leq -1/2$ (see Fig.\,3). The expressions for the three positive roots $y_j$ and the angle $\psi$ simplify accordingly to
\begin{equation}y_1= {1\over 3}(1+\cos\psi)\,,\qquad y_{2,3} = {1\over 6}\big(2\pm \sqrt{3} \sin\psi - \cos\psi\big)\,, \qquad \psi = {1\over 3} \arccos(-10-18\zeta)\,. \end{equation}
As a result the matrix exponential function for the exceptional (real) Lie algebra $\mathbf{g}_2$ and corresponding compact Lie group $G_2$ is solved analytically by the formula in eq.(32), setting $\xi=1$ and $\eta_7 = -(1+2\zeta)/12$, and implementing the linear relations written in eq.(39) into $V$ and $\hat v$. 

The $\mathbf{g}_2$-constraint $\xi=1$ implies for the three roots the additional relation $y_1^2+y_2^2+y_3^2=1/2$. This translates into a constraint on the angles $\varphi_j = V\!\sqrt{y_j}$:
 \begin{equation}2(\varphi_1^4 +\varphi_2^4 +\varphi_3^4 )-(\varphi_1^2 +\varphi_2^2 +\varphi_3^2)^2 = ( \varphi_1+\varphi_2 +\varphi_3) ( \varphi_1+\varphi_2 -\varphi_3) ( \varphi_1-\varphi_2 +\varphi_3) ( \varphi_1-\varphi_2 -\varphi_3)=0\,,
 \end{equation}
 which in view of the chosen ordering is solved  by $\varphi_1=\varphi_2 +\varphi_3$. Consequently, the trace of $G_2$-matrices in $SO(7)$ is given by
 \begin{equation}
  \text{tr}R_{G_2}(\vec v\,) =1+ 2 \cos(\varphi_2 +\varphi_3) +2 \cos\varphi_2+2 \cos\varphi_3= 8 \cos{\varphi_2 \over 2}\,\cos{\varphi_3 \over 2} \,\cos{\varphi_2+\varphi_3 \over 2}-1\,, \end{equation}
 and the two independent angles $\varphi_2, \varphi_3$ parametrize a two-dimensional maximal torus  $SO(2)\times SO(2) = \mathbf{S}^1 \times \mathbf{S}^1$ in $G_2$.  Stated differently, by passing to the subgroup $G_2\subset SO(7)$  the angle for the rotation in the $x_1x_2$-plane is fixed to the sum of the rotation angles in the $x_3x_4$- and $x_5x_6$-planes, $\varphi_1=\varphi_2 +\varphi_3$. 
  
   \section{SO(8)}
Elements of the 28-dimensional Lie algebra $so(8)$ are antisymmetric $8\times 8$ matrices of the form
\begin{equation}\vec J\!\cdot\! \vec v  =  \left( \begin{array} {cccccccc} 0 & v_1& v_2&v_3&v_4&v_5&v_6& v_7
\\ -v_1& 0& v_8&v_9 &v_{10}& v_{11}& v_{12}& v_{13}\\ -v_2& -v_7& 0 &v_{14}&v_{15}&v_{16} &v_{17} &v_{18}\\ -v_3& -v_9& -v_{14}& 0&v_{19}& v_{20}& v_{21} & v_{22}\\ -v_4& -v_{10}& -v_{15}&-v_{19}& 0&v_{23}& v_{24}&v_{25}\\ -v_5& -v_{11}& -v_{16}&-v_{20}& -v_{23}& 0& v_{26}&v_{27}\\ -v_6& -v_{12}& -v_{17}&-v_{21}& -v_{24}& - v_{26}&0  &v_{28}\\ -v_7& -v_{13}& -v_{18}&-v_{22}& -v_{25}& - v_{27}&-v_{28} & 0\cr \end{array}\!\right), 
\end{equation} 
with the corresponding norm $ V=\sqrt{v_1^2+\dots +v_{28}^2}$ of the 28-component real vector $\vec v$. The characteristic polynomial \cite{koecher} of $\vec J\!\cdot \!\hat v$ is of degree eight and it reads
 \begin{equation} P_8(x) = x^8+x^6+x^4\Big({1\over 2}-{\xi\over 4}\Big) +x^2\Big({1-\zeta\over 6}-{\xi\over 4}\Big)+\eta \,,
 \end{equation}
 with the invariants $\xi = \text{tr}(\vec J\!\cdot \!\hat v)^4,\, \zeta = \text{tr}(\vec J\!\cdot \!\hat v)^6$ and $\eta = \text{det}(\vec J\!\cdot \!\hat v)$ forming its  coefficients.  By setting  $y=-x^2$ one is lead  to a quartic polynomial
 \begin{equation} \tilde P_4(y) = y^4-y^3+y^2\Big({1\over 2}-{\xi\over 4}\Big)+ y\Big({\zeta-1 \over 6}+{\xi\over 4}\Big) +\eta\,,
 \end{equation}
 whose four roots $y_1, y_2, y_3, y_4\geq 0$ all have to be positive, since those of $P_8(x)=0$ are the purely imaginary eigenvalues of the antihermitean $8\times 8$ matrix $\vec J\!\cdot \!\hat v$.
The determination of the four roots $y_j$ proceeds via three auxiliary quantitites $\Theta_1,  \Theta_2, \Theta_3$ in the following way \cite{algebra}
\begin{equation} y_{1,2} = {1\over 4} \Big[1+ \sqrt{\Theta_1} \pm \big( \sqrt{\Theta_2} + \sqrt{\Theta_3}\,\big) \Big] \,, \qquad   y_{3,4} = {1\over 4} \Big[1 -\sqrt{\Theta_1} \pm \big( \sqrt{\Theta_2} - \sqrt{\Theta_3}\,\big) \Big] \,, \end{equation} 
where the product of the three square-roots must fulfil the condition $\sqrt{\Theta_1}\sqrt{\Theta_2} \sqrt{\Theta_3} = 1/3-\xi- 4\zeta/3$. This means that if the right hand side is negative, $3\xi+4\zeta>1$, one square-root must be chosen as negative. The three positive $\Theta$-values are the roots of the cubic resolvent polynomial 
\begin{equation} R_3(\Theta) =\Theta^3+\Theta^2(1-2\xi) +\Theta\Big( {5\over 3}-64\eta -4\xi +\xi^2 -{8\zeta\over 3}\Big) -{1\over 9} \big( 3\xi+ 4\zeta -1\big)^2 \,.\end{equation}
The solution of $R_3(\Theta)=0$ belongs again to the irreducible case and is perfomed with the trigonometric ansatz
 \begin{equation} \Theta = {1\over 3} \big( 2\xi - 1 + 2\sqrt{192\eta +8\zeta+8\xi+\xi^2-4}  \cos \psi \big) \,.\end{equation}
 The determining equation for $\cos 3\psi$  yields for the trisected angle $\psi$ the result 
  \begin{equation}\psi = {1\over 3}\arccos{8(1-36\eta -3\zeta+3\zeta^2)-\xi^3+42 \xi^2+12\xi(5\zeta +48\eta -3)\over (192\eta +8\zeta+8\xi+\xi^2-4)^{3/2}  }\,,\end{equation} and the three auxiliary quantities $\Theta_1, \Theta_2, \Theta_3$ follow by evaluating eq.(51) at the angles $\psi$ and $\psi\pm 2\pi/3$. The necessary conditions that the radicand in eq.(52) is positive, and that the numerator is smaller in magnitude that the denominator follow from considering the local minimum and maximum located at $\Theta_\pm =( 2\xi-1 \pm   \sqrt{192\eta +8\zeta+8\xi+\xi^2-4}\,)/3\geq 0$ together with $R_3(\Theta_+)R_3(\Theta_-)\leq0$. One should remark here that the cubic polynomial $R_3(\Theta)$ has actually the same form as $\tilde P_3(y)$ shown in Fig.\,1.

Based on the Cayley-Hamilton relation $\Sigma^8 = -\Sigma^6 +\Sigma^4( \xi-2)/4+\Sigma^2[\xi/4+(\zeta-1)/6] -\eta \, \mathbf{1}$  one constructs the $8\times 8$ iteration matrix as
 \begin{equation}M_8  =  \left( \begin{array} {cccccccc}0 & 0& 0&0 &0& 0& 0 &-\eta \\ 1& 0 & 0& 0&0 &0& 0&0 \\ 0& 1 & 0& 0&0 &0& 0 &\xi/4\!+\!(\zeta\!-\!1)/6\\  0& 0& 1 & 0& 0&0 &0& 0\\
0&  0& 0& 1& 0& 0& 0 &\xi/4\!-\!1/2\\ 0& 0 &0& 0& 1 & 0 & 0& 0 
 \\0&  0 &0 & 0& 0& 1& 0&-1\\ 0& 0 &0& 0& 0& 0& 1& 0\cr \end{array}\!\right), \end{equation} 
with which one can solve $\exp(V M_8)$ and multiply with $(1,0,0,0,0,0,0,0)$. 
This way one derives the following quasi-analytical formula for the matrix exponential function $so(8)\to SO(8)$: 
\begin{align} R_8(\vec v\,)=\exp(\vec J \!\cdot \!\vec v\,) =\sum_{j=1}^4 &\Big[ 4y_j^3-3 y_j^2+y_j\Big(1\!-\!{\xi\over 2}\Big) +{\xi\over 4}+{\zeta\!-\!1\over 6}\Big]^{-1}  
\Big\{ \cos\big(V\!\sqrt{y_j}\big) \mathbf{1}+ {1\over \sqrt{y_j}} \sin\big(V\!\sqrt{y_j}\big)  \vec J \!\cdot \!\hat v\Big\} \nonumber \\ & \times\Big\{ -{\eta \over y_j} \mathbf{1} +\Big( {\xi\over 4}-{1\over 2}+y_j-y_j^2\Big) (\vec J \!\cdot \!\hat v)^2+(y_j-1) (\vec J \!\cdot \!\hat v)^4 -(\vec J \!\cdot \!\hat v)^6 \Big\} \,,
\end{align} 
where the sum goes over the four positive roots of the quartic equation $\tilde P_4(y)=0$. Note that the denominator of the prefactor is the derivative $\tilde P_4'(y_j)$, and one gets a remarkable factorization for the terms with (up to seventh) powers of $\vec J \!\cdot \!\hat v$. The trace of the orthogonal matrix  $R_8(\vec v\,)$ comes out as 
\begin{equation} \text{tr} R_8(\vec v\,) =2 \sum_{j=1}^4 \cos\big(V\!\sqrt{y_j}\big)\,,\end{equation}
where the simplification is achieved by eliminating for each summand $\eta$ in favor of the root $y_j$ through the equation $\tilde P_4(y_j)=0$. 
The four angles $\varphi_j = V\!\sqrt{y_j}$ specify to which element on a four-dimensional maximal torus $SO(2)\times SO(2)\times SO(2)\times SO(2) = \mathbf{S}^1 \times \mathbf{S}^1\times \mathbf{S}^1\times \mathbf{S}^1$ 
a given $SO(8)$-matrix $R_8(\vec v\,)$ is related by conjugation     \cite{liegroup}.

   \section{SO(9)}
Elements of the 36-dimensional Lie algebra $so(9)$ are antisymmetric $9\times 9$ matrices of the form
\begin{equation}\vec J\!\cdot\! \vec v  =  \left( \begin{array} {ccccccccc} 
0 & v_1& v_2&v_3&v_4&v_5&v_6& v_7&v_8\\ 
-v_1& 0& v_9&v_{10} &v_{11}& v_{12}& v_{13}& v_{14} & v_{15}\\  
-v_2&  -v_9& 0 &v_{16}&v_{17}&v_{18} &v_{19} &v_{20}& v_{21}\\ 
-v_3& -v_{10}& -v_{16}& 0&v_{22}& v_{23}& v_{24} & v_{25} &v_{26}\\ 
-v_4& -v_{11}& -v_{17}&-v_{22}& 0&v_{27}& v_{28}&v_{29}& v_{30} \\ 
-v_5& -v_{12}& -v_{18}&-v_{23}& -v_{27}& 0& v_{31}&v_{32}&v_{33}\\ 
-v_6& -v_{13}& -v_{19}&-v_{24}& -v_{28}& - v_{31}&0  &v_{34} & v_{35}\\ 
-v_7& -v_{14}& -v_{20}&-v_{25}& -v_{29}& - v_{32}&-v_{34} & 0&v_{36}\\ 
-v_8& -v_{15}& -v_{21}&-v_{26}& -v_{30}& - v_{33}&-v_{35} & -v_{36}&0   \cr \end{array}\!\right), 
\end{equation} 
with the corresponding norm $ V=\sqrt{v_1^2+\dots +v_{36}^2}$ of the 36-component real vector $\vec v$. The characteristic polynomial \cite{koecher} of $\vec J\!\cdot \!\hat v$ is of degree nine and it reads
 \begin{equation} P_9(x) = x^9+x^7+x^5\Big({1\over 2}-{\xi\over 4}\Big) +x^3\Big({1-\zeta\over 6}-{\xi\over 4}\Big)+x\Big( {1\over 24}-{\xi+\chi\over 8} +{\xi^2\over 32}-{\zeta \over 6} \Big) \,, \end{equation}
with a new invariant, $\chi = \text{tr}(\vec J\!\cdot \!\hat v)^8$, the trace of the eighth matrix power. Besides the trivial root $x=0$ of $P_9(x)$, the other purely imaginary ones are found via the substitution $y=-x^2$ from the quartic polynomial
 \begin{equation} \tilde P_4(y) = y^4-y^3+y^2\Big({1\over 2}-{\xi\over 4}\Big) +y\Big({\zeta-1 \over 6}+{\xi\over 4}\Big)+  {1\over 24}-{\xi+\chi\over 8} +{\xi^2\over 32}-{\zeta \over 6} \,, \end{equation}
 which after identification of the constant term with $\eta$ is identical to $\tilde P_4(y) $ in eq.(48) pertinent to the case of $so(8)$. Therefore, when working with the auxiliary parameter 
 \begin{equation} \eta_9= {1\over 24}-{\xi+\chi\over 8} +{\xi^2\over 32}-{\zeta \over 6} \,, \end{equation}
 the construction of the four positive roots $y_1,y_2,y_3,y_4$ can be copied from the previous section by making merely the substitution  $\eta \to \eta_9$.
 The denominator of the arc-cosine function in eq.(52) becomes this way
 \begin{equation} 192\eta_9 +8\zeta+8\xi+\xi^2-4= 7\xi^2-16\xi-24(\zeta+\chi)+4\,, \end{equation}
while the numerator polynomial turns into
 \begin{eqnarray}
&&  8(1-36\eta_9 -3\zeta+3\zeta^2)-\xi^3+42 \xi^2+12\xi(5\zeta +48\eta_9 -3) \nonumber \\ && = \xi^2(17\xi-39) +12\xi(2-3\zeta-6\chi) +12( 2\zeta+2\zeta^2+3\chi) -4 \,. \end{eqnarray}

Based on the Cayley-Hamilton relation $\Sigma^9 = -\Sigma^7 +\Sigma^5( \xi-2)/4+\Sigma^3[\xi/4+(\zeta-1)/6] -\Sigma \,\eta_9 $  one constructs the $9\times 9$ iteration matrix as 
 \begin{equation}M_9  =  \left( \begin{array} {ccccccccc}0 & 0& 0&0 &0& 0& 0 & 0&0 \\ 1&0 & 0& 0&0 &0& 0& 0 &-\eta_9 \\ 0& 1& 0 & 0& 0&0 &0& 0&0 \\ 0& 0& 1 & 0& 0&0 &0& 0 &\xi/4\!+\!(\zeta\!-\!1)/6\\ 0&  0& 0& 1 & 0& 0&0 &0& 0\\0& 
0&  0& 0& 1& 0& 0& 0 &\xi/4\!-\!1/2\\ 0& 0& 0 &0& 0& 1 & 0 & 0& 0 
 \\0& 0&  0 &0 & 0& 0& 1& 0&-1\\ 0& 0& 0 &0& 0& 0& 0& 1& 0\cr \end{array}\!\right), \end{equation} 
with which one can solve $\exp(V M_9)$ and multiply with $(1,0,0,0,0,0,0,0,0)$. 
This way one derives the following quasi-analytical formula for the matrix exponential function $so(9)\to SO(9)$: 

\begin{eqnarray} R_9(\vec v\,)=\exp(\vec J \!\cdot \!\vec v\,) &=& \mathbf{1}+{1\over \eta_9} \Big\{ \Big( {1-\zeta\over 6}-{\xi \over 4}\Big) (\vec J \!\cdot \!\hat v)^2+ \Big({1\over 2} -{\xi\over 4}\Big) (\vec J \!\cdot \!\hat v)^4 +(\vec J \!\cdot \!\hat v)^6 +(\vec J \!\cdot \!\hat v)^8\Big\}  \nonumber \\ && +2 \sum_{j=1}^4
\Big[9y_j^4- 7y_j^3+{5\over 4}y_j^2(2-\xi)+{ y_j\over 4}(3\xi+2\zeta-2)  +\eta_9\Big]^{-1} \nonumber \\ &&  \qquad \quad \times 
\Big\{ {\eta_9 \over y_j} \mathbf{1} +\Big({1\over 2}-{\xi \over 4} -y_j+y_j^2\Big) (\vec J \!\cdot \!\hat v)^2  +(1-y_j) (\vec J \!\cdot \!\hat v)^4 +(\vec J \!\cdot \!\hat v)^6 \Big\}\nonumber \\ &&  \qquad \quad \times 
\Big\{ \cos\big(V\!\sqrt{y_j}\big) (\vec J \!\cdot \!\hat v)^2 -\sqrt{y_j} \sin\big(V\!\sqrt{y_j}\big)  \vec J \!\cdot \!\hat v \Big\} \,,
\end{eqnarray} 
The calculation of the  trace of the orthogonal matrix  $R_9(\vec v\,)$ gives the result
\begin{equation} \text{tr} R_9(\vec v\,) =1+ 2 \sum_{j=1}^4 \cos\big(V\!\sqrt{y_j}\big)\,, \end{equation}
where one first uses $\chi=1/3-\xi+\xi^2/4-4\zeta/3- 8 \eta_9$ and then eliminates $\eta_9$ in favor of the root $y_j$ via the equation $\tilde P_4(y_j)=0$. 
The four angles $\varphi_j = V\!\sqrt{y_j}$ specify to which element on  a four-dimensional maximal torus $SO(2)\times SO(2)\times SO(2)\times SO(2) = \mathbf{S}^1 \times \mathbf{S}^1\times \mathbf{S}^1\times \mathbf{S}^1$ a given $SO(9)$-matrix  $R_9(\vec v\,)$ is related by conjugation \cite{liegroup}. By convention the independent rotations take place in the $x_1x_2$-plane ,  $x_3x_4$-plane, $x_5x_6$-plane, and  $x_7x_8$-plane of nine-dimensional space. 

It is hoped that the given quasi-analytical formulas will be  useful to generate the full manifold of special orthogonal $SO(n)$-matrices and (exceptional) $G_2$-matrices in various applications.

\subsection*{Acknowledgement} This work  has been supported in part by DFG (Project-ID 196253076 - TRR 110) and NSFC.

\end{document}